\documentclass[twocolumn, showpacs,amsmath,amssymb,nofootinbib]{revtex4}

\usepackage{graphicx}
\usepackage{dcolumn}
\usepackage{bm}
\usepackage{epsfig}
\usepackage{color}

\usepackage{color}
\usepackage{hyperref}

\hypersetup{
    colorlinks=true,
    linkcolor=red,
    citecolor=blue,
}

\def\fun#1#2{\lower3.6pt\vbox{\baselineskip0pt\lineskip.9pt
  \ialign{$\mathsurround=0pt#1\hfil##\hfil$\crcr#2\crcr\sim\crcr}}}
\def\simgt{\mathrel{\lower0.6ex\hbox{$\buildrel {\textstyle >}
 \over {\scriptstyle \sim}$}}}
\def\simlt{\mathrel{\lower0.6ex\hbox{$\buildrel {\textstyle <}
 \over {\scriptstyle \sim}$}}}
\def\mK{{\mathcal K}}
\def\bea{\begin{eqnarray}}
\def\eea{\end{eqnarray}}
\def\be{\begin{equation}}
\def\ee{\end{equation}}
\def\tr{{\rm tr}\,}

\input epsf

\def\be{\begin{equation}}
\def\ee{\end{equation}}
\def\ba{\begin{eqnarray}}
\def\ea{\end{eqnarray}}

\begin{document}

\preprint{}

\title{Analytic solutions in non-linear massive gravity}

\author{Kazuya Koyama$^{1}$\email{Kazuya.Koyama@port.ac.uk}, Gustavo Niz$^{1}$, Gianmassimo Tasinato$^{1}$}

\bigskip

\affiliation{$^1$Institute of Cosmology \& Gravitation, University of Portsmouth, Dennis Sciama Building, Portsmouth, PO1 3FX, United Kingdom \\}


\begin{abstract}
We study spherically symmetric solutions in a covariant massive gravity model, which is a candidate for a ghost-free non-linear completion of the Fierz-Pauli theory. There is a branch of solutions that exhibits the Vainshtein mechanism, recovering General Relativity below a Vainshtein radius given by $(r_g m^2)^{1/3}$, where $m$ is the graviton mass and $r_g$ is the Schwarzschild radius of a matter source. Another branch of exact solutions exists, corresponding to Schwarzschild-de Sitter spacetimes where the curvature scale of de Sitter space is proportional to the mass squared of the graviton.
\end{abstract}



\maketitle
{\bf Introduction}:
It is a fundamental question whether there exists a consistent covariant theory for massive gravity, where the graviton acquires a mass and leads to a large distance modification of General Relativity (GR). The quest for massive gravity dates back to the work by Fierz and Pauli (FP) in 1939 \cite{Fierz:1939ix}. They considered a mass term for linear gravitational perturbations, which explicitly breaks the gauge invariance of GR. As a result, there exist five degrees of freedom in the graviton, instead of the two found in GR. There have been intensive studies on what happens  going beyond the linearised theory. In 1972, Boulware and Deser (BD) found that, at the non-linear level, there appears a sixth mode in the graviton that becomes a ghost 
  in the FP model \cite{Boulware:1973my}. This problem was reexamined using the effective theory approach \cite{ArkaniHamed:2002sp}, where additional (St\"uckelberg) fields were introduced to restore the gauge invariance, and whose scalar part represents the helicity-0 mode of the graviton. In the FP model, the scalar acquires non-linear interaction terms that contain more than two time derivatives, signaling the existence of the ghost.

The St\"uckelberg approach also sheds light on the other puzzle in the FP gravity: if one linearises the system, the solutions in the FP theory do not reduce to GR solutions in the massless limit. This is known as the van Dam, Veltman, Zakharov (vDVZ) discontinuity \cite{vanDam:1970vg, Zakharov:1970cc}. However, in this massless limit the scalar mode becomes strongly coupled and one cannot linearise the system. Therefore, due to strong coupling, the scalar interaction is shielded and GR can be recovered. This is known as the Vainshtein mechanism \cite{Vainshtein:1972sx}.
The strong coupling scale in the FP model is identified as $\Lambda_5 =(m^4 M_{pl})^{1/5}$ where $M_{pl}$ is the Planck scale and $m$ is the graviton mass. This scale is tightly connected with the non-linear interactions of the scalar mode that contain more than two time derivatives. In the decoupling limit, where $m \rightarrow 0$ and $M_{pl} \to \infty$, while the strong coupling scale $\Lambda_5$ is kept fixed, one obtains an effective theory for the scalar mode, where it is possible to study the consistency of the theory in more detail.

Until recently, it was believed that there is no consistent way to extend the FP model \cite{Deffayet:2005ys, Creminelli:2005qk}
to get a ghost free model at all orders. A breakthrough came with a 5D braneworld model known as Dvali-Gabadadze-Porrati (DGP) model \cite{Dvali:2000hr}.
 In this model  there appears a continuous tower of massive gravitons
 from a four dimensional perspective.
  The non-linear interactions of the helicity-0 mode of massive gravitons contain no more than two derivatives, which is crucial to avoid the BD ghost. Due to this fact, the strong coupling scale in this theory is given by $\Lambda_3 =(m^2 M_{pl})^{1/3}$ instead of $\Lambda_5$, where $m=r_c^{-1}$ and $r_c$ is a cross-over scale between 5D and 4D gravity \cite{Luty:2003vm, Nicolis:2004qq}. Further studies have considered more general non-linear interactions which contain no more than two derivatives. In 4D, only a finite number of terms satisfy this condition; these are dubbed Galileon terms because of a symmetry under field transformations of the form $\partial_{\mu} \pi \to \partial_{\mu} \pi + c_{\mu}$ \cite{Nicolis:2008in}. Ref.~\cite{deRham:2010ik} constructed the extension of the FP theory that gives the Galileon terms in the decoupling limit, by choosing the correct parameters in the Lagrangian up to quintic order in perturbations. Ref.~\cite{claudia} proposed a covariant non-linear action that automatically ensures this property to all orders, which we will discuss below.

A remaining crucial question is whether this property, holding
 in the decoupling limit,  is sufficient to ensure the absence of the BD ghost or not. In Ref.~\cite{claudia}, it was shown that there is no BD ghost in the decoupling limit to all orders in perturbation theory, but only up to and including quartic order away from this limit. However, it is very hard to show the absence of the BD ghost at all orders if one starts from Minkowski and studies non-linear interactions perturbatively. Therefore, it is important to obtain non-perturbative background solutions in this theory, and study fluctuations around them. Moreover, it is interesting to find solutions in this covariant non-linear theory, that can describe features of the observed universe. These are the topics of the present work.

{\bf Covariant non-linear massive gravity}:
We first construct the action for generalised FP model based on Ref.~\cite{deRham:2010ik, claudia}.
We define the tensor $H_{\mu \nu}$ as a covariantization of the metric perturbations:
\be
g_{\mu \nu} = \eta_{\mu \nu} +h_{\mu\nu}\,\equiv\,H_{\mu \nu}+\eta_{\alpha \beta}
\partial_\mu \phi^\alpha \partial_\nu \phi^\beta.
\ee
The St\"uckelberg fields $\phi^\alpha\,=\,\left(x^\alpha-\pi^\alpha\right)$
transform as scalars, while $\eta_{\alpha \beta}$ corresponds to a non-dynamical
background metric that is needed to define the potential, which is assumed to be the
Minkowski metric. The covariant tensor $H_{\mu \nu}$ can then be expanded as
\bea
H_{\mu \nu}&=&h_{\mu \nu}+\eta_{\beta \nu}\partial_\mu \pi^\beta+\eta_{\alpha \mu}\partial_\nu
\pi^\alpha-
\eta_{\alpha \beta} \partial_\mu \pi^\alpha \partial_\nu \pi^\beta, \nonumber\\
&\equiv& h_{\mu \nu}-{\mathcal Q}_{\mu\nu},
\eea
and under the coordinate transformation $x^{\mu} \to x^{\mu} + \xi^{\mu}$,
$\pi^{\mu}$ transforms as
\be\label{pi_trans}
\pi^{\mu} \to \pi^{\mu} + \xi^{\mu}.
\ee
Before proceeding with the massive gravity theory, indices are raised/lowered, from now on, with the dynamical metric $g_{\mu\nu}$; for example $H^\mu_{\,\,\nu}\,=\,g^{\mu \rho} H_{\rho \nu}$.

We define a new tensor ${\mathcal K}_{\,\,\mu}^\nu$ as
\be
{\mathcal K}_{\mu}^{\ \nu} \equiv \delta_{\mu}^{\ \nu}-\left(\sqrt{g^{-1} \left[g-H\right]}\right)_{\mu}^{\ \nu},
\ee
where the square root is formally understood as
$\sqrt{A}_{\mu}^{\ \alpha}\sqrt{A}_{\alpha}^{\ \nu}=A_{\mu}^{\ \nu}$.
This allows us to represent the complete potential for gravitational interactions as
\be\label{genlag}
{\cal L} = \frac{M_{Pl}^2}{2}\,\sqrt{-g}\left( R- m^2{\cal U}\right), \;\; {\cal U } = \left[ \tr(\mK^2)-(\tr \mK)^2\right].
\ee
By expanding the potential in $H_{\mu \nu}$, we get an infinite sum
of interaction terms for $H_{\mu \nu}$, with the FP term at the
  lowest order.

In order to study exact solutions associated with the previous
Lagrangian, it is convenient to express $\mathcal K$ in terms
of matrices, namely
\be
\mK = {\mathbb I}-\sqrt{g^{-1} \left[\eta+\mathcal Q\right]},
\ee
where ${\mathbb I}$ denotes the identity matrix, and we have used $H_{\mu \nu}= g_{\mu \nu}-(\eta_{\mu \nu} +Q_{\mu \nu})$.
The potential in four dimensions then
reads
\bea\label{genepotexp}
{\cal U }&=&  \tr g^{-1}  \left[\eta+\mathcal Q\right] -12\nonumber\\
&+& \tr \sqrt{g^{-1}
 \left[\eta+\mathcal Q\right]
}\left(
6-  \tr \sqrt{g^{-1} \left[\eta+\mathcal Q\right]} \right).
\eea
In general, the task is to calculate
the trace of $\sqrt{g^{-1}  \left[\eta+\mathcal Q\right]}$.  Given that $g^{-1}  \left[\eta+\mathcal Q\right]$ is a square matrix, 
the Schur decomposition theorem ensures that it can be expressed as
\be
g^{-1} \left[\eta+\mathcal Q\right]\,=\,{\cal T} \,D \,{\cal T}^{-1},
\ee
for some unitary matrix ${\cal T} $, and an upper triangular matrix $D$. The diagonal entries of $D$ are 
the eigenvalues of $g^{-1} \left[\eta+\mathcal Q\right]$, and we call these eigenvalues $\lambda_1,\dots,\lambda_4$.
Then, since $\sqrt{g^{-1}  \left[\eta+\mathcal Q\right]}\,=\,{\cal T} \,\sqrt{D} \,{\cal T}^{-1}$,
one can express the traces in the formulae above, in terms of eigenvalues, as
\bea
\tr{g^{-1}  \left[\eta+\mathcal Q\right]}&=&\sum_i \lambda_i, \nonumber \\
\tr{\sqrt{g^{-1}  \left[\eta+\mathcal Q\right]}}&=&\sum_i \sqrt{\lambda_i}.
\eea
Plugging these expressions into the potential, Eq. (\ref{genepotexp}),
we find the following expression for ${\cal U}$
\bea\label{genepotexp2}
{\cal U }&=&\sum_i \lambda_i+
\Big(\sum_j
\sqrt{\lambda_j}
\Big)\Big(
6-  \sum_i
\sqrt{\lambda_i}\Big)-12.
\eea

{\bf Spherically  symmetric configurations}:
We now focus on
 analyzing properties of
spherically
symmetric configurations in this set-up.
We start by considering static
configurations in the unitary gauge, $\pi^{\mu}=0$ (see Ref.~\cite{Damour:2002gp} for spherical symmetric solutions in the FP theory).
The most general form of the metric respecting spherical symmetry is
\be\label{genmetr}
d s^2\,=\,-C(r) \,d t^2+A(r)\, d r^2 +2 D(r)\, dt dr+B(r) d \Omega^2,
\ee
where $d \Omega^2 = d \theta^2 + \sin^2 \theta d \phi^2$.
We choose to write the non-dynamical flat metric as $ds^2 = -dt^2 + dr^2 + r^2 d \Omega^2$.
Notice that in GR one can set $B(r)=r^2$ by a coordinate transformation,
but this is not possible here, since we have already fixed the gauge completely.
In order to simplify the analysis, it is convenient to define the combination
$\Delta(r)\,=\,A(r) C(r)+D^2(r)$. We plug  the previous metric into the Einstein equations
\be\label{einstein}
G_{\mu \nu}=T^{{\cal U}}_{\mu \nu},
\ee
where the energy momentum tensor from the potential ${\cal U}$ of Eq. (\ref{genepotexp2})
is defined as $T^{{\cal U}}_{\mu \nu}\,=\frac{m^2}{\sqrt{-g}}\,\frac{ \delta \sqrt{-g} {\cal U}}{\delta g^{\mu \nu}}$.
The Einstein tensor $G_{\mu\nu}$ satisfies the identity $D(r)\, G_{tt}+C(r)\,G_{tr}\,=\,0$, which implies
the algebraic constraint
 \bea
 0&=& D(r)\, T^{{\cal U}}_{tt}+C(r)\,T^{{\cal U}}_{tr} \nonumber\\
 &=& m^2\,\frac{
 D(r)\,
 \left(2 r -3 \sqrt{B(r)}\right)
 \,\sqrt{\Delta(r)}
 }{\sqrt{B(r)}\,\left( A(r)+C(r) +2\sqrt{\Delta(r)}\right)^{1/2}}\label{eqbranches}.
 \eea

The previous condition can be satisfied in two ways, which lead to {\it two} different branches of solutions.
We can either set $D(r)=0$, and focus  on diagonal metrics, or alternatively, set $B(r)\,=\,4\,r^2/9$.
The fact that there are two branches of solutions indicate that, unlike in GR where Birkhoff's theorem holds,
there is no uniqueness theorem for spherically symmetric solutions in this theory. We analyze the two branches in turn.



{\bf Diagonal-metric branch - asymptotically flat solutions}:
The case of diagonal metrics, $D(r)=0$ in Eq. (\ref{genmetr}), leads to equations which in general cannot be solved analytically. Thus, we will analyze them perturbatively, 
showing that they lead to asymptotically flat solutions (see Ref.~\cite{us} for a more detailed discussion on this branch of solutions).
We expand the functions $A, C$ and $B$ as
\be
A(r) = \frac{1}{1 + f}, \;\; C(r)= (1 + n)^2 , \;\;  B(r)= \frac{r^2} {(1+h)^2},
\ee
and truncate the field equations to first order in $n$, $f$ and $h$. It is more convenient to introduce a new radial coordinate $\rho = \sqrt{B(r)}$, so that the linearised metric is expressed as
\be
ds^2 = - (1 + 2  n) dt^2 + (1 - \tilde{f}) d\rho^2
+ \rho^2 d \Omega^2,
\ee
where $\tilde{f} =  f - 2  h - 2 \rho  h'$ and a prime denotes a derivative with respect to $\rho$. 
In this new coordinate, the solutions for $n$ and $\tilde{f}$ are then given by
\bea
2 n &=& - \frac{8 G M}{3 \rho} e^{- m \rho}, \nonumber\\
\tilde{f} &=& -\frac{4 G M}{3 \rho} (1 + m \rho) e^{- m \rho},
\label{linsol}
\eea
where we fix the integration constant so that $M$ is the mass of a point particle at the origin and $8 \pi G = M_{pl}^{-2}$. See the right plot of Fig.~\ref{plot} for the general behaviour of these solutions.
Notice that these configurations, as anticipated, are asymptotically flat and exhibit
the vDVZ discontinuity, {\it i.e.} they do not agree with the GR solutions ($2n_{GR}=\tilde{f}_{GR}=-2GM/\rho$) in the limit $m \to 0$. However, in order to understand what really happens in this limit, one should take into account the non-linear behaviour of $h$. Let us consider scales below the Compton wavelength $m \rho \ll 1$, and at the same time ignore higher order terms in $G M$.
Under these approximations, the equations of motion can still be truncated to linear
order in $\tilde{f}$ and $n$, but since $h$ is not necessarily small, we keep all non-linear terms in $h$. Then we obtain the following equations
\bea
2 \rho  n' &=& \frac{2 G M}{\rho} - (m \rho)^2 h, \nonumber\\
\tilde{f} &=& - 2 \frac{G M}{\rho} - (m \rho)^2 (h - h^2), \nonumber\\
\frac{G M}{\rho} &=& - (m \rho)^2
\left(\frac{3}{2} h - 3 h^2 + h^3 \right).
\label{solh}
\eea
We should stress that these are exact equations in the limit $m \rho \ll 1$, $G M/\rho \ll 1$,
{\it i.e.} there are no higher order corrections in $h$. For large radial $\rho$ values, one can linearise the equations in $h$, recovering the solution in Eqs.~(\ref{linsol}), to first order in $m \rho$. On the other hand, the Vainshtein mechanism applies, and below the so-called Vainshtein radius, $\rho_V = (G M m^{-2})^{1/3}$,
$h$ becomes larger than one and the non-linear terms in $h$ become important, recovering GR close to a matter source. Actually, for $\rho \ll \rho_V$ the solution for $h$ is simply given by $|h| = \rho_V/\rho \gg 1$.
The latter solution for $h$ and Eq. (\ref{solh}) imply
\bea
2 \rho n' &=& \frac{ 2 G M}{\rho} \left(1 + \frac{1}{2} \left(\frac{\rho}{\rho_V}
\right)^2 \right),\nonumber \\
\tilde{f} &=& - \frac{ 2 G M}{\rho} \left(1 -\frac{1}{2} \left(\frac{\rho}{\rho_V}
\right) \right).\label{solfbv}
\eea
Therefore, corrections to the GR solutions are indeed small for $\rho<\rho_V$, as shown in the left plot of Fig.~{\ref{plot}.

The Vainshtein mechanism becomes also transparent in a non-unitary gauge. Indeed by performing the coordinate transformation $\rho = \sqrt{B(r)}$, we excite the $\rho$ component of the St\"uckelberg field (see Eq.~({\ref{pi_trans})), $ \pi^{\rho} = - \rho h$. Thus the strong coupling nature of $h$ is encoded in $\pi^{\rho}$ in this coordinate. It is possible to construct an effective theory for this St\"uckelberg field in the so-called decoupling limit \cite{deRham:2010ik}: first we introduce a scalar so that $\pi_{\mu} = \partial_{\mu} \pi/\Lambda_3^3$, where
$\Lambda_3^3 = m^2 M_{pl}$.
Then the covariantization of metric perturbations $H_{\mu \nu}$ is written as
\be
H_{\mu \nu} = h_{\mu \nu} + \frac{2}{M_{pl} m^2} \Pi_{\mu \nu}
- \frac{1}{M_{pl}^2 m^4} \Pi^2_{\mu \nu},
\label{H}
\ee
where $\Pi_{\mu \nu} = \partial_{\mu} \partial_{\nu} \pi$ and
$\Pi^2_{\mu \nu} = \Pi_{\mu \alpha} \Pi^{\alpha}_{\nu}$.
Formally, the decoupling limit is achieved by taking $m \to 0$ and $M_{pl} \to \infty$, but keeping $\Lambda_3$ fixed. By substituting Eq.~(\ref{H}) into the action,
one can show that the kinetic terms of $\pi$ become total derivatives and a mixing
 appears
between $h_{\mu \nu}$ and $\pi$, which can be diagonalised using the definition
\be
h_{\mu \nu} = \hat{h}_{\mu \nu} +\frac{\pi}{M_{pl}} \eta_{\mu \nu}
- \frac{1}{\Lambda_3^3 M_{pl}} \partial_{\mu} \pi \partial_{\nu} \pi.
\label{metric}
\ee
The Lagrangian is then written as 
\bea
{\cal L} &=& {\cal L}_{\rm GR} (\hat{h}_{\mu \nu})
+ \frac{3}{2} \pi \Box \pi - \frac{3}{2 \Lambda_3^3}
(\partial \pi)^2 \Box \pi \nonumber\\
&+&
\frac{1}{2 \Lambda_3^6} (\partial \pi)^2
([\Pi^2] - [\Pi]^2) \nonumber\\
&+& \frac{5}{2 \Lambda_3^9} (\partial \pi)^2
([\Pi]^3 - 3 [\Pi] [\Pi^2]+2 [\Pi^3]),
\eea
where $[\Pi]$ = $\Pi^{\mu}_{\mu}$, $[\Pi^2] = \Pi^{\mu \nu} \Pi_{\mu \nu}$, $[\Pi^3]=\Pi^{\mu \nu} \Pi_{\nu \alpha} \Pi^{\alpha}_{\mu}$ and
${\cal L}_{\rm GR}$ is the linearised Einstein-Hilbert action for $\hat{h}_{\mu \nu}$.
The terms containing the scalar field $\pi$ are known as Galileon terms, which give rise to the second order
differential equations, as explained in the Introduction. For the
 spherically symmetric case, the equation of motion for $\pi$ simplifies to \cite{Nicolis:2008in, Wyman:2011mp}
\be
3 \left(\frac{\pi'}{\rho}\right)
+ \frac{6}{\Lambda_3^3} \left(\frac{\pi'}{\rho}\right)^2
+  \frac{2}{\Lambda_3^6} \left(\frac{\pi'}{\rho}\right)^3
= \frac{M}{4 \pi\,M_{pl}\, \rho^3},
\label{pieq}
\ee
where the integration constant is again chosen so that $M$ is a mass of a particle at the origin. Using the relation between $\pi$ and $h$,
$h = -\pi'/\left(m^2 M_{pl} \rho\right)$,
we can show that the solutions for $\tilde{f}$, $n$ and $h$ given by Eq.~(\ref{solh}) agree with
the solutions Eq.~(\ref{metric}) and Eq.~(\ref{pieq}).

We have shown that the weak field solutions for the metric Eq.~(\ref{genmetr}) with $D(r)=0$ have three phases (see Fig.~\ref{plot}). On the largest scales, $ m^{-1} \ll \rho  $, beyond the Compton wavelength, the gravitational interactions decay exponentially due to the mass of graviton: see Eq.~(\ref{linsol}) and region 3 in Fig.~\ref{plot}.
In the intermediate region $\rho_V < \rho < m^{-1}$, we obtain the $1/r$ gravitational potential but Newton's constant is rescaled $G \to 4 G/3$. Moreover, the post-Newtonian parameter $\gamma$ is $\gamma=\tilde{f}/(2n)=(1/2)(1+m\rho)$, which reduces to $\gamma=1/2$ in the $m\rightarrow 0$ limit, instead of $\gamma=1$ of GR, showing the vDVZ discontinuity (see region 2 in Fig.~\ref{plot}). Finally, below the Vainshtein radius $\rho<\rho_V$, the GR solution is recovered due to the strong coupling of the $\pi$ mode (see Eq.~(\ref{solfbv}) and region 1 in Fig.~\ref{plot}).
   This background solution provides us with a testing ground for the BD ghost. Instead of expanding the action in $H_{\mu \nu}$ around the Minkowski spacetime perturbatively, one can study {\it linear} perturbations around this non-perturbative solution using the complete
     potential Eq.~(\ref{genepotexp}). In order to obtain the fully non-linear solution, a numerical approach is necessary. In the next section, we consider the second branch of
     solutions for this theory, which can instead be obtained analytically.

\begin{figure}[htp!]
{\centering
\resizebox*{3.3in}{1.8in}
{\includegraphics{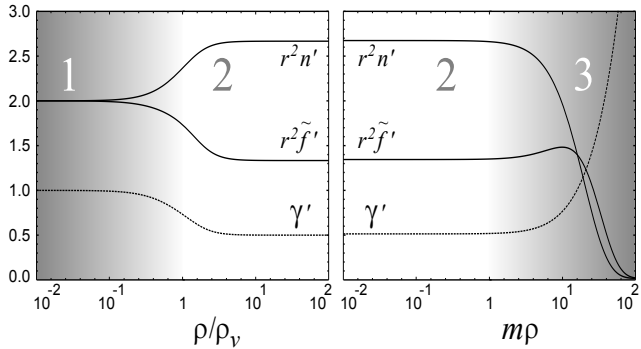}} }
\caption{Numerical solution for $\partial_r\tilde{f}=\tilde{f}'$, $\partial_r n=n'$, and the quotient $\gamma'\equiv\tilde{f'}/2n'$ around the Vainshtein radius $\rho_v$ (left) and the Compton Wavelength $\rho\sim 1/m$ (right). Region 1 (2) shows how GR solutions are (not) recovered inside (outside) the Vainshtein radius $\rho_V$. Region 3 shows the asymptotic decay of the linear solutions (Eq.~(\ref{linsol})). Here,  $GM=1$.}
\label{plot}
\end{figure}

{\bf
 Non-diagonal-metric branch - Schwarzschild de Sitter solutions}:
Next, we analyze the second branch of vacuum solutions that 
solve Eq. (\ref{eqbranches}), where
$B(r) =4\,r^2/9$.
Interestingly, this branch leads to asymptotically de Sitter configurations.
There is another identity $C(r) T^{{\cal U}}_{rr} + A(r) T^{{\cal U}}_{tt}=0$, 
which leads to the condition
\be
\Delta(r)  \,=\,A(r) C(r)+D^2(r) \,\equiv\,\Delta_0\,=\,\mathrm{const}.
\ee
The remaining Einstein equations provide the following unique solution
(see Ref.~\cite{us} for detailed derivations)
%
\bea
A(r) &=& \frac{9 \Delta_0}{4} (p(r) + \alpha +1),\;\;\;\;\;
B(r) =\frac49 r^2,\label{coeffdss}\\
C(r) &=&\frac{9 \Delta_0}{4}(1 - p(r)), \;\;
D(r) = \frac{9}{4} \Delta_0 \sqrt{p(r)(p(r)+\alpha)}, \nonumber
\eea
where
\be
p(r)=\frac{2 \mu}{r} + \frac{m^2 r^2}{9}, \;\;
\alpha =\frac{16}{81\,\Delta_0}-1,
\ee
with  arbitrary $\mu$ and $\Delta_0$.
This solution is
similar
to that in \cite{Salam:1976as}, up to numerical factors.
Notice that  this configuration
depends on {\it two} integration constants.
 A sufficient condition to ensure that $D(r)$ is real, is to choose $\mu\geq 0$ and $0<\sqrt{\Delta_0} \leq 4/9$.
The form of
metric coefficients as in  Eq.(\ref{coeffdss}) do  not allow a
manifest comparison with de Sitter spacetime, since we have already chosen the unitary gauge and cannot do a further coordinate transformation without exciting components of $\pi^\mu$.
However, if we allow for a
vector $\pi^\mu$ of the form $\pi^\mu\,=\,\left( \pi_0(r),0,0,0\right)$,
the metric can be rewritten in a diagonal form as
\be\label{diagmetf}
d s^2\,=\,-C(r) \,d t^2+\tilde{A}(r)\, d r^2  + B(r) d \Omega^2.
\ee
Then we can write down the action in terms of $C, \tilde{A}$, $B$ and $\pi_0$,
considering them as fields. It is possible to show that the following configuration solves the corresponding
equations of motion
\be
\tilde{A}(r) = \frac{4}{9} \frac{1}{1 - p(r)}, \;\; \pi_0'(r) = - \frac{\sqrt{p(r)(p(r)+\alpha)}}{1-p(r)},
\ee
while  $C(r)$ and $B(r)$ are the same as in Eq.~(\ref{coeffdss}).
The resulting metric has then a manifestly de Sitter-Schwarzschild form by making a time rescaling
$t \rightarrow  (4/9 \Delta_0^{1/2}) t$. However, we should note that this time rescaling cannot be done without
introducing an additional time dependent contribution to $\pi_0$.
As expected, the metric in Eq. (\ref{diagmetf}) can be obtained by making the following transformation of the time coordinate
$d \tilde t\,\equiv\, d t + \pi_0' d r$  to the metric (\ref{genmetr});
this produces a non-zero time component of $\pi^{\mu}$, that does not vanish even in the $m\rightarrow 0$ limit for any allowed value of $\Delta_0$. There are two integration constants, $\mu$ and $\Delta_0$, in this solution. In GR, $\mu$ corresponds to the mass of a source at the origin but a careful analysis including a matter source
is necessary to fully understand the role of these integration constants.
Note that there is an apparent singularity at the horizon $p(r)=1$, both for the metric and $\pi_0$.

We can further make a coordinate transformation at the expense of exciting further components of $\pi^{\mu}$. For example, by setting $\mu=0$ and
making the following coordinate transformations
$ t = F_t(\tau, \rho), \;\; r = F_r(\tau, \rho)$
with
\bea
F_t(\tau, \rho) &=& \frac{4}{3 \Delta_0^{1/2} m} {\rm arctanh}
\left(
\frac{ \sinh \left(\frac{m \tau}{2} \right) + \frac{m^2 \rho^2}{8} e^{m \tau/2}}
{\cosh \left(\frac{m \tau}{2} \right) - \frac{m^2 \rho^2}{8} e^{m \tau/2}}
\right), \nonumber\\
F_r(\tau, \rho) &=& \frac{3}{2} \rho e^{m \tau/2},
\eea
the metric becomes that of flat slicing of de Sitter,
\be
ds^2 = - d \tau^2 + e^{m \tau} (d\rho^2 + \rho^2 d \Omega^2),
\ee
where the Hubble parameter is given by $m/2$.
The St\"uckelberg fields $\pi^{\mu}$ are now given by
$
\pi^{\mu} = (\pi^{\tau}(\tau, \rho), \pi^{\rho}(\tau, \rho),0,0),$
$
\pi^{\tau} = \pi_0 +\tau- F_t(\tau, \rho), \;\;
\pi^{\rho}=  \rho- F_r(\tau, \rho)\,.
$
This is an interesting solution in which the acceleration of the universe is determined by the graviton mass and the Hubble parameter is given by $m/2$. For $\Delta_0=16/81$, this solution reduces to the ``self-accelerating" solution obtained in the decoupling limit in Ref.~\cite{deRham:2010tw}.

{\bf Conclusions}:
The solutions obtained in the non-linear covariant massive gravity are remarkably similar to those in the DGP braneworld model including the existence of the ``self-accelerating" de Sitter solution without  cosmological constant \cite{Deffayet:2000uy} although there are differences in detail. There are a number of important issues. Firstly, we should confirm that there is no BD ghost in this theory by studying perturbations around the non-perturbative solution obtained in this letter. In the DGP model, the self-accelerating solution suffers from a ghost instability \cite{Luty:2003vm, Nicolis:2004qq, Koyama:2005tx}, which is related to the ghost in the FP theory on a de Sitter background. It is crucial to study the stability of the de Sitter solution in this model. In fact Ref.~\cite{deRham:2010tw} showed that there exists a ghost in this self-accelerating background in the decoupling limit for a particular value of the second integration constant $\Delta_0$. They argue that this ghost can be cured by adding higher order corrections in ${\cal K}$ to the potential. Our formalism is ready to be applied to this extended model.  However, we believe a more complete analysis of perturbations about our exact solution is needed. Once these issues are clarified, the massive gravity model presented here provides an interesting playground to study large distance modifications of general relativity.

\acknowledgments
We would like to thank Misao Sasaki for useful comments, and Timothy Clemson for reading the manuscript.
KK and GN are supported by the ERC. KK is supported by STFC grant ST/H002774/1 and GT is supported by STFC advanced fellowship. KK acknowledges support from the Leverhulme trust and RCUK.

\end{document}